# X-ray Metrology of an Array of Active Edge Pixel Sensors for Use at Synchrotron Light Sources

R. Plackett, K. Arndt, D. Bortoletto, I. Horswell, G. Lockwood, I. Shipsey, N. Tartoni, S. Williams

**Abstract**

We report on the production of an array of active edge silicon sensors as a prototype of a large array. Four Medipix3RX.1 chips were bump bonded to four single chip sized Advacam active edge n-on-n sensors. These detectors were then mounted into a 2 by 2 array and tested on B16 at Diamond Light Source with an x-ray beam spot of 2um. The results from these tests, compared with optical metrology give confidence that these sensors are sensitive to the physical edge of the sensor, with only a modest loss of efficiency in the final two rows of pixels. We present the efficiency maps recorded with the microfocus beam and a sample powder diffraction measurement. These results give confidence that this sensor technology can be used in much larger arrays of detectors at synchrotron light sources.

1) **Introduction**

Hybrid Pixel Detectors are now used ubiquitously at synchrotron lights sources. Their high dynamic range, fast readout, high quantum efficiency and low services requirements make them increasingly popular. A limitation with this technology is the inability to construct large continuously sensitive arrays of detector elements. Currently, the sensor of the hybrid is typically bounded by a 500um inactive guard ring structure to keep the leakage current to a manageable level. To minimise the insensitive area in tiled detector arrays large area sensors are bump-bonded to multiple small area readout ASICs. These large sensor-multi-ASIC modules are difficult to produce and suffer yield issues with one badly bonded ASIC resulting in module with up to 16 ASICs being discarded. An alternative that has recently become available is active edge, or 'edgeless' sensors, where the guard ring is replaced with a trench etching process that uses implants on the edge of the silicon. This approach allows the single-sensor-multi-ASIC detector modules to be simply replaced with an array of sensor-single-ASIC modules that are placed in very close proximity to each other. Maneuski et al[1] describe work characterising the performance of these sensors and show very promising results. To demonstrate that there is no significant loss in efficiency between an active edge sensor and a current multi-ASIC device, an array of four edgeless Medipix3[2] single chip modules were mounted on a Diamond Light Source quad printed circuit board that was designed to hold a quad multi ASIC array. This was tested on the B16 beamline at Diamond by raster scanning a micro-focus x-ray beam across the inner corners of the array to determine the fraction of x-rays lost in this region. Although this work was carried out with Medipix3 ASICs, it is applicable to all hybrid detectors of this type.

2) **Sensor Preparation**

The active edge sensor array to be tested was assembled in the Oxford Physics Micostructure Detector (OPMD) lab from components supplied by Quantum Detectors Ltd and Advacam Oy. The readout ASICs were Medipix3RX.1 chips designed by the Medipix3 collaboration and supplied with a Merlin readout system as part of development kit by Quantum Detectors specifically for this project.

The active edge sensors were 200um thick, 55um pixelated devices of an n-on-n type, supplied by Advacam Oy, who also performed the flip chip bonding between the ASIC and sensor.

The individual active edge assemblies were glued one at a time to the Merlin Quad readout PCB using a thermally conductive, RTV silicone glue that would allow rework. The second, third and fourth chips were positioned so they were in contact with the chips that had previously been glued down, allowing a minimal gap to be achieved. It is acknowledged that this manual assembly method could be improved with the use of vacuum clamping to restrain the assemblies whilst the glue cures. The array was then wire bonded to the readout PCB and the interconnect fidelity tested. The final assembled detector array is shown in Figure 1.

A water-cooled aluminum block was mounted behind the detector head to keep it at a constant temperature of 15C. Although the system is capable of running without active temperature stabilisation, for these tests it was preferable to exclude this potential performance variable from the measurements.

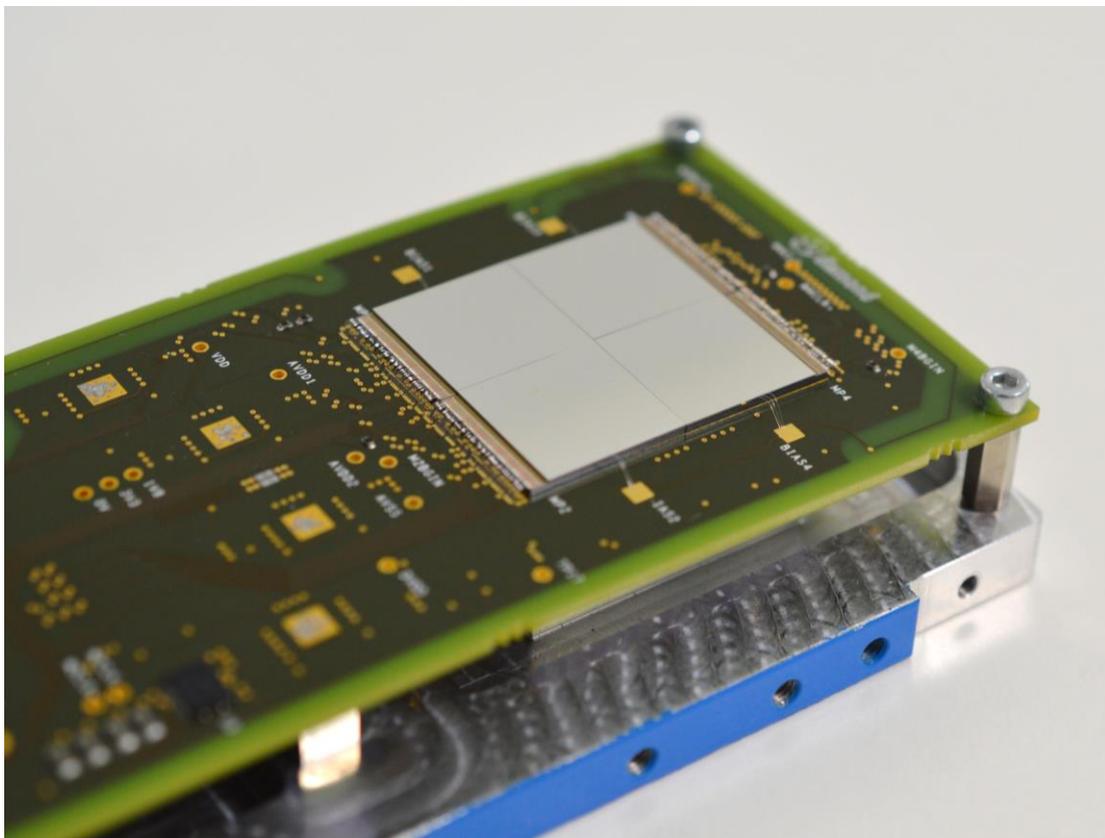

*Figure 1: The active edge sensor array after gluing to the carrier PCB and wirebonding. The Gaps between the sensors are visible as are the sensor bias wires.*

3) **Optical Metrology**

Metrology was performed with a Keyence VHX 5000 calibrated digital microscope and OGP CNC 500 optical metrology system to ascertain the physical alignment of the assemblies in the array. The smallest gap achieved between the lower two modules was measured as 16.5um with the calibrated microscope, and 18 +/-2um with the OGP metrology system. The other three gaps were measured as 56um, 56um and 57um with the OGP metrology system. A calibrated microscope image is shown in Figure 2 below.

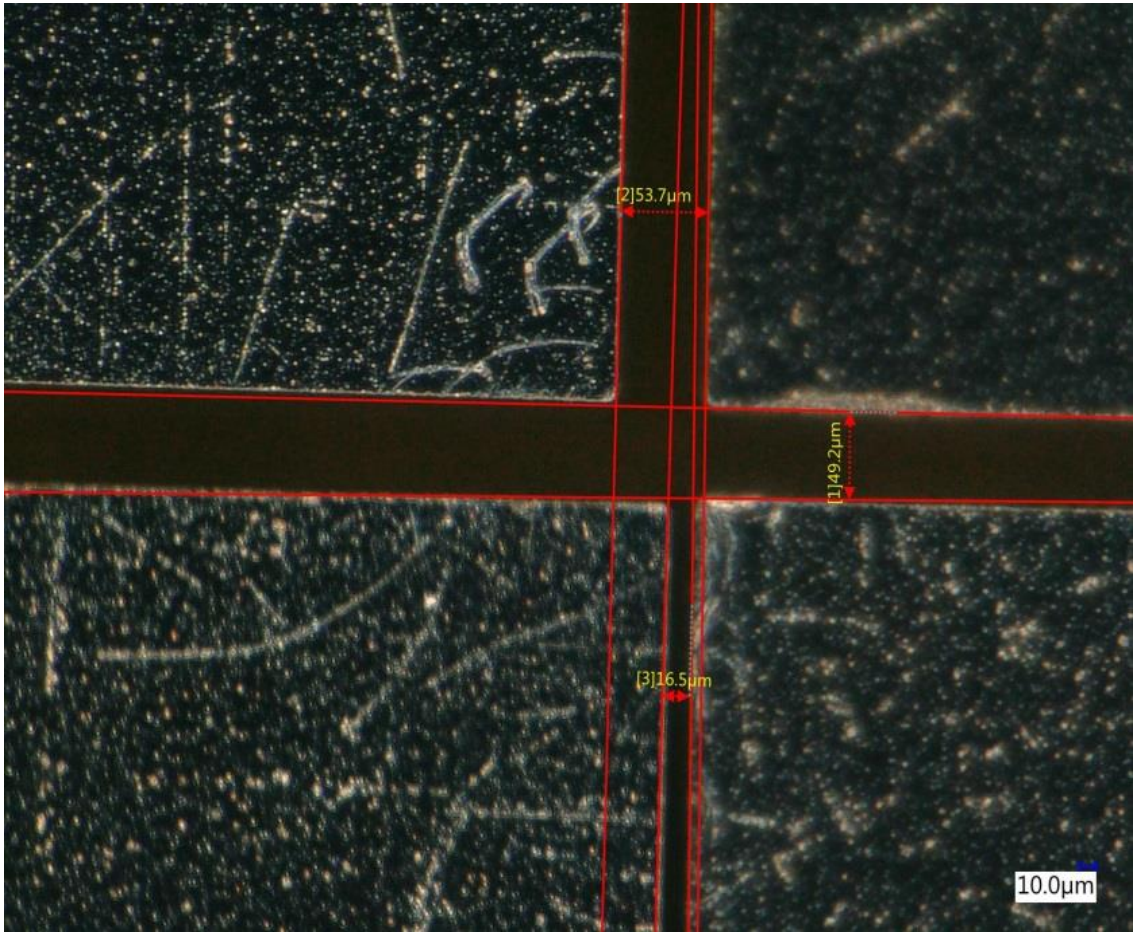

*Figure 2: An image of the central four corners of the edgeless array at 500x magnification, taken with the calibrated microscope and showing the fitted measurement lines.*

### 4) Beamline Integration

The sensor array was connected to the Merlin readout system that was integrated into the beamline software DAQ systems to ease data taking. This allowed the detector to be fully controlled by the beamline GDA/EPICS system which increased the speed and level of automation with which the data was taken.

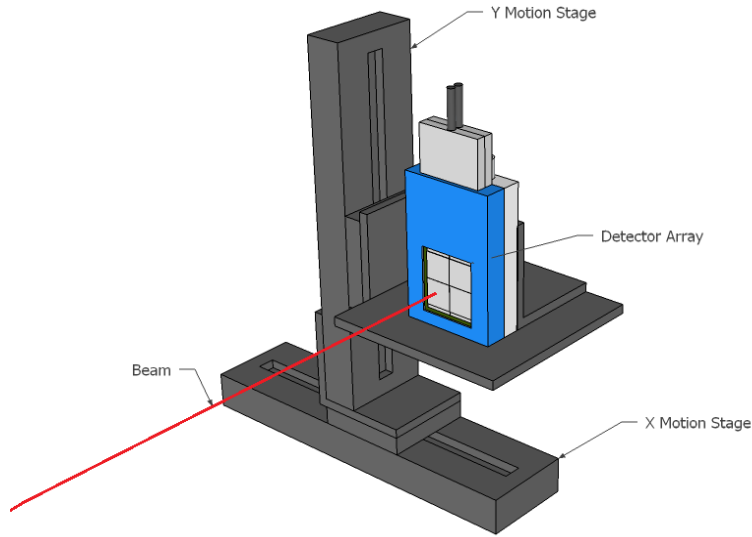

*Figure 3: The setup of the detector array in the micro-focus beam, showing the alignment of the system relative to the beam and the two motion stages that were used to perform the raster scans.*

5) **Raster Scans**

A Compound Reflective Lens (CRL) X-ray mirror arrangement was used to focus the monochromatic beam at B16 to a spot of 2um diameter (FWHM). The beam energy was set to 15.7keV, which was partly determined by the requirement for a small spot size, and partly to match the fluorescent energy of Zirconium. The detector was mounted onto a pair of motion stages, as shown in Figure 3, in the focal plane and moved across the focal point in both orthogonal dimensions in 5 +/-1um steps. Data was acquired for 0.5 seconds at each step to build up a sub-pixel composite image of the efficiency and pixel field structure of the region of interest. This scan was performed over a significant region of several square millimeters in the array's centre.

The extent of these scans are shown in Figures 4 and 5 which are maps of the highest individual pixel photon count on the matrix for the raster position plotted. This effectively maps the response of the individual pixels, but does not take into account the increase in efficiency due to counting in multiple pixels in the pixel edge and corner regions. Figure 6 shows the central region, corrected for this effect by summing the counts in the neighbouring pixels to the highest responding pixel. The regions of lower efficiency in the corners are due to the system being set with a slightly high threshold level compared with its ultimate gain. The threshold is set above the nominal 50% of operating energy where the charge sharing effects would cancel out. Here efficiency is simply defined as a fraction of the maximum expected counts, measured when the beam hits the central region of a pixel, where the system is expected to be fully efficient and record every photon.

Line scans of the individual pixel responses are shown in Figures 7 & 8 for two positions and, as with Figures 4 through 6, the minimal gap between chips 1 and 2 is confirmed to be 15um. This close correspondence with the optical metrology confirms that there is no discernible region of the active edge sensor that does not have some sensitivity, and that the reduced sensitivity is limited to the outer two rows of pixels.

These two rows of pixels exhibit a larger than usual collection area, which can be easily seen in the line scans and the 2D raster scans by comparing them with pixels further from the edge of the sensors. Their much greater physical extent, and lower absolute collection efficiency, stems from the distortions in the field lines inherent in the design of the active edges. The line scans in Figures 7 and 8 reveal that the efficiency remains at 0.6 of nominal until the edge of the sensor in these two rows of pixels. In particular, the loss of efficiency in these pixels is unimportant as it can be compensated for by a flat field correction as it is stable and does not change over time. Of greater concern is the ambiguity introduced by the sensitivity of the pixels overlapping, again clearly shown in the line scans (see Figures 7 and 8). Due to this, spatial information from these pixels should be treated with some caution in the same way that data from the larger pixels in the traditional arrays are treated, either interpolating or otherwise identifying them to identify the loss in granularity.

This distortion effect strengthens at the corners with the second row from the edge in particular suffering from rotational distortions for several pixels. The third and fourth pixel rows appear normal in the raster scan in Figure 4, however the per-pixel line scans in Figures 7 and 8 reveal that they both suffer some spreading. This spreading, and the distortion of the second row, leads to the 'hot edge' effect that has been previously reported[1] and can be seen in the flat field image in Figure 10. The hot edge effect is the apparent over efficiency of pixels in the $2^{nd}$ $3^{rd}$ and $4^{th}$ rows away from the edge and is caused by their effective size being increased by the field line distortions, giving higher counts when under a uniform illumination.

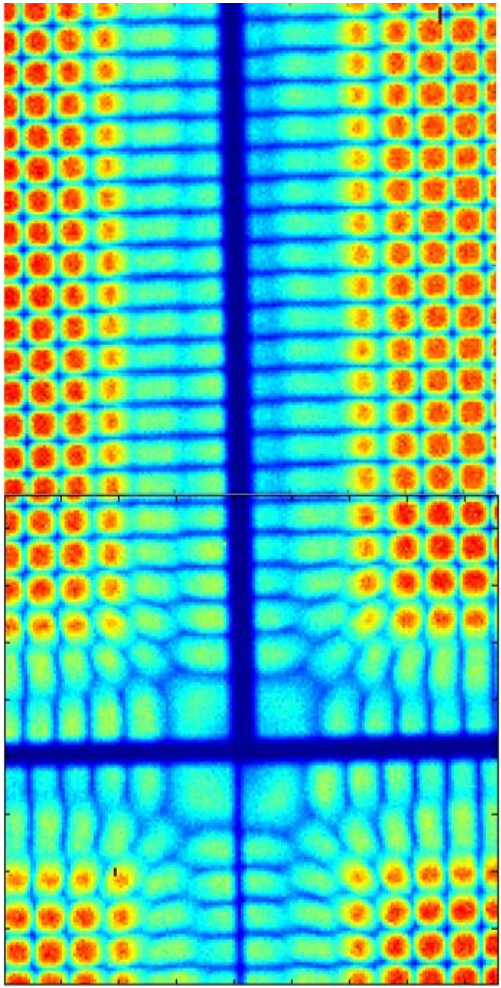

*Figure 4: A composite image showing the result of two raster scans over the central region of the array. The scan step size is 5 um and the colour axis denotes the photons received in the most active pixel in that image. The 15um and 55um gaps between the devices can be clearly seen as can a number of interesting field line effects on the corner pixels.*

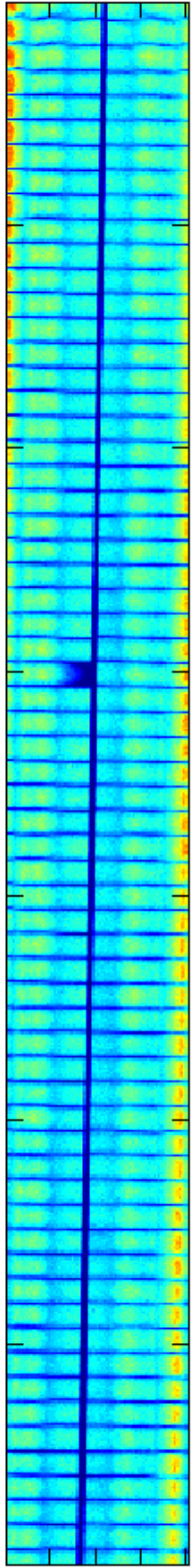

*Figure 5: A continuation of the scans shown in Figure 2 with the same step and acquisition parameters, but different scan geometry. This further explores the smaller gap and demonstrates the quality of the overall alignment compared to the pixel size.*

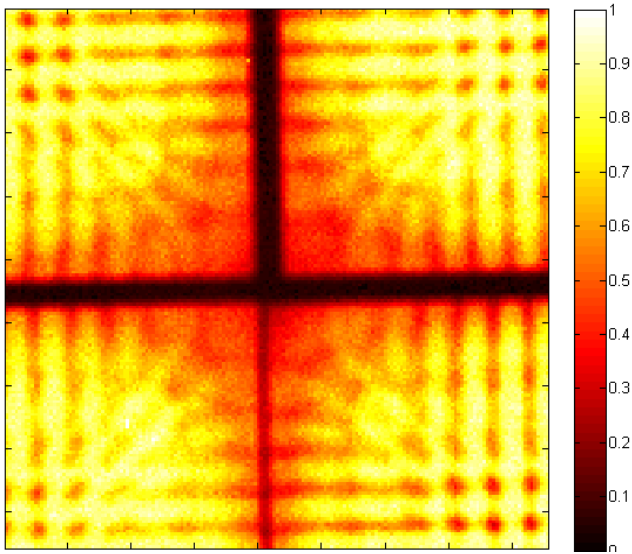

*Figure 6: The 'true' efficiency of the central raster scan. This image shows the normalised efficiency on the colour axis to counteract the loss effect caused by charge sharing. The combined counts are the sum of the most active pixel and all its neighbours.*

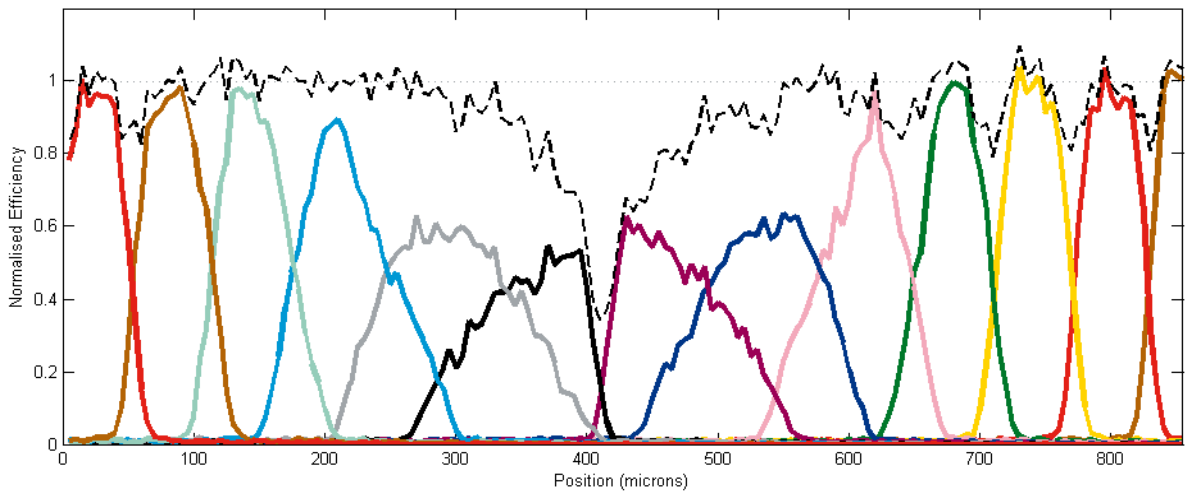

*Figure 7: A one-dimensional section from the data shown in Figure 6 across the boundary with the 16.5um gap. The individual pixel responses are shown in solid colours and the summed data, correcting for charge sharing plotted in Figure 5, is shown as the heavy dashed line.*

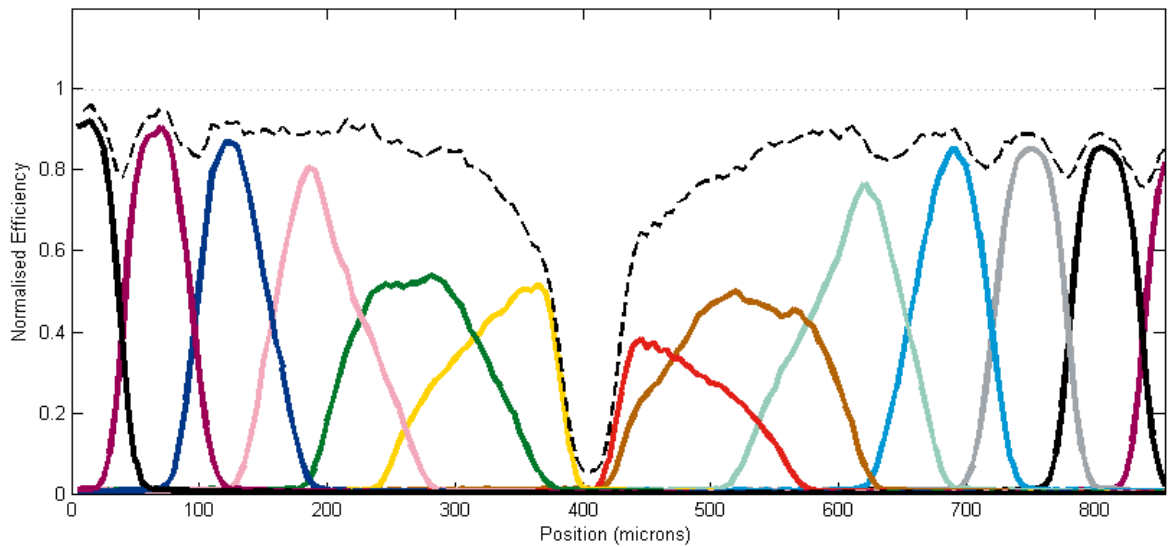

*Figure 8: A one-dimensional section from the data shown in Figure 5 across the boundary with the 57um gap. The individual pixel responses are shown in solid colours and the summed data, correcting for charge sharing plotted in Figure 6, is shown as the heavy dashed line. The individual pixel data are shown with higher photon statistics than in Figure 7 as the repeated matrix allowed the data from the same point in 15 rows to be averaged.*

### 6) Powder Diffraction Measurement and Flat Field

To demonstrate the utility of the sensor array, a standard measurement of a powder diffraction ring was performed. This was separated into two parts, a flat field image to correct for the 'hot edge' effect and a measurement of a silicon powder sample.

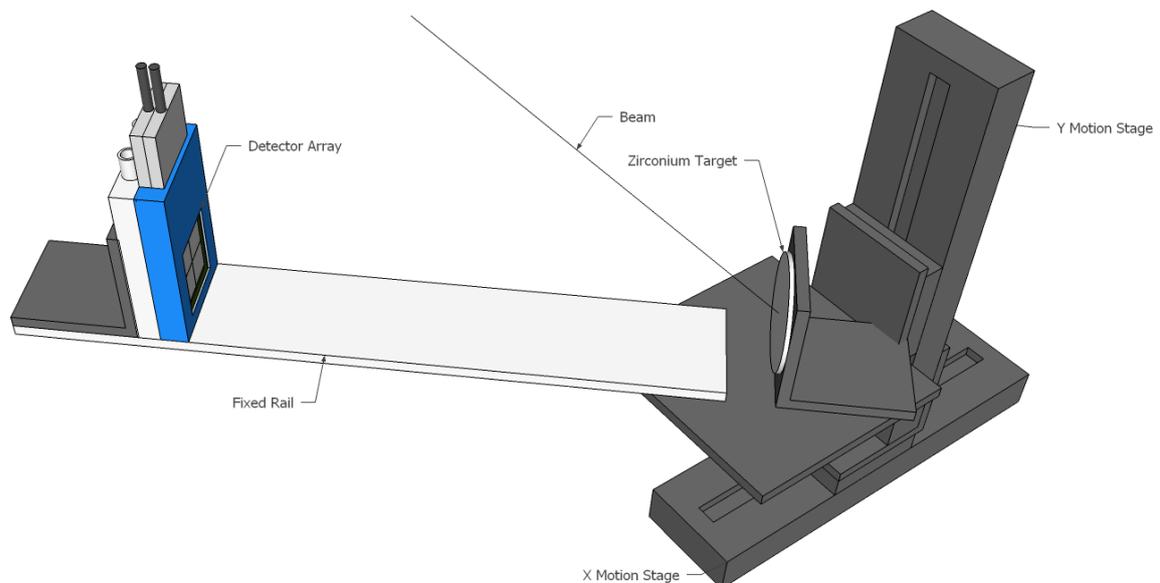

*Figure 9: The experimental setup for the flat field illumination of the detector array. The Zirconium target is mounted at 45 degrees to the incident beam and the detector array is mounted parallel to the target at a distance of 300mm.*

To achieve a flat field a zirconium target was used as a source of uniformly scattered x-rays. The detector was set 300mm away from, parallel to the target at a 45 degree angle to the incident beam, as shown in Figure 9. To achieve a faster measurement, the CRL mirror was removed and the target was illuminated with the beam set to a monochromatic energy of 18.2keV and a spot size of 5mm. This energy, which is above the appropriate absorption line of Zirconium, was used to allow fluorescence at 15.7keV. A long duration acquisition was taken to ensure that all pixels counted more than 10,000 counts to reduce the statistical uncertainty. The flat field image recorded is shown in Figure 10 and clearly shows the hot edge region caused by the extended pixels and the lower efficiency of the edge-most pixels. This image has not been corrected to account for the gap or larger area of the edge pixels and displays all the pixels in the matrix as if they were the same size.

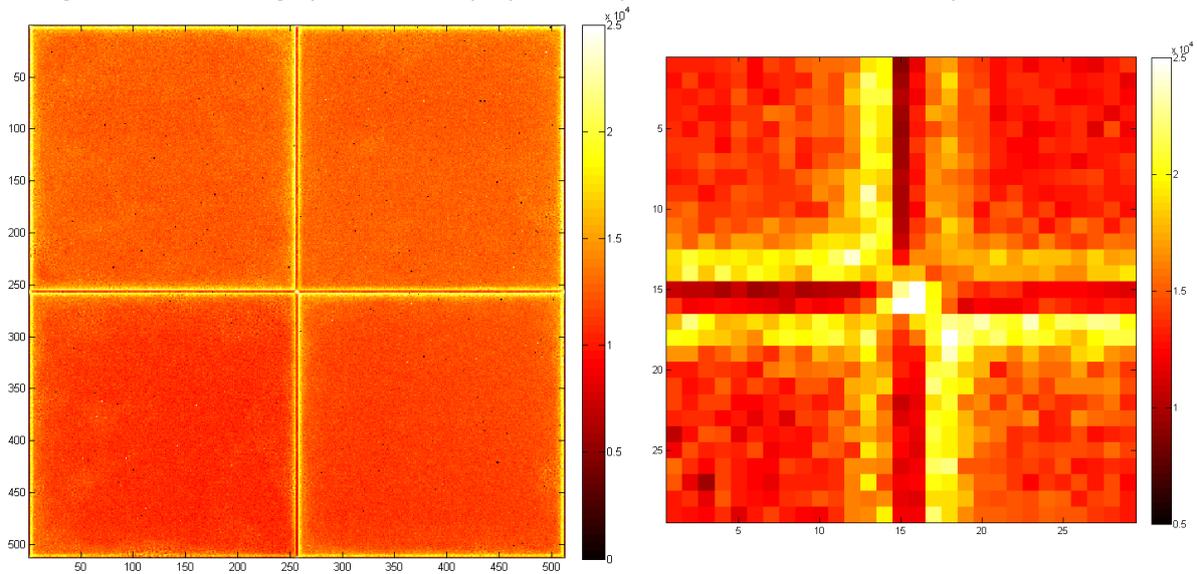

*Figure 10: (left) The flat field image recorded using the setup shown in Figure 8. The 'hot edge' effects of the edge pixels are clearly visible as are the lower efficiency of the last edge pixels. This image is unprocessed and shows all the pixels to be the same size. (right) A zoomed in view of the central area showing details of the hot edge effect.*

The system was then returned to the original configuration shown in Figure 3 and a silicon powder sample was mounted 30mm in front of, and slightly above, the detector array. The powder sample was illuminated with the micro-focus beam and the data recorded as shown in Figure 11 and Figure 12. Also shown in these figures is the image once the flat field correction has been applied, with the majority of the hot edge and inefficient edge pixels corrected for.

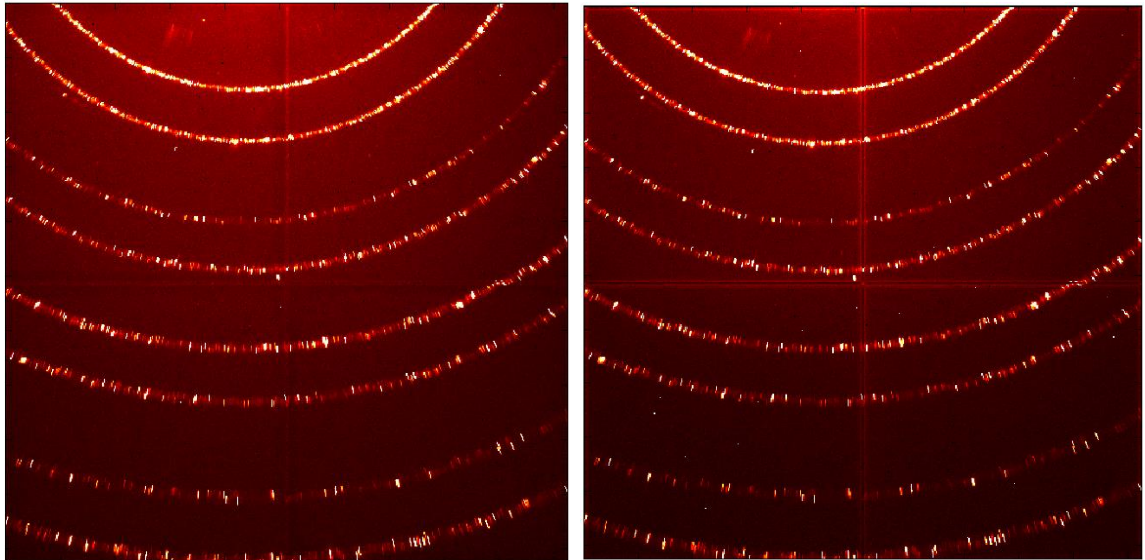

*Figure 11: Silicon Powder rings with (left) and without (right) the flat field correction applied. The hot edge effects are very largely reduced and the low efficiency edge pixels compensated for. The rings recorded match the expected powder diffraction pattern for silicon.*

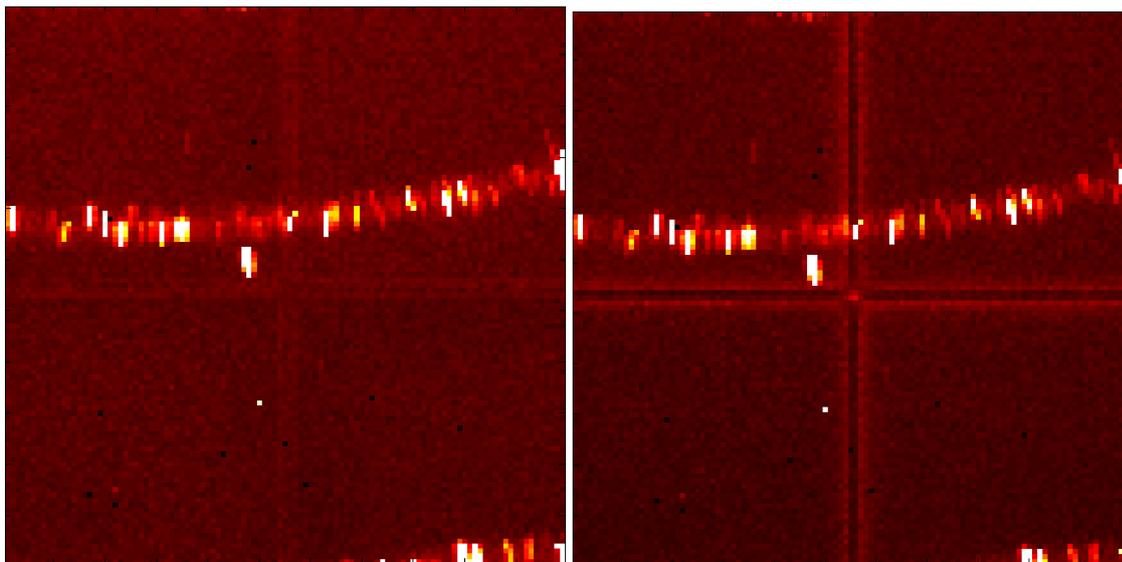

*Figure 12: An image of the centre of the powder ring image before (left) and after (right) the flat field correction is applied showing the effect of the correction at the internal corner of the sensors.*

### 7) Conclusions and Future Work

By comparing the raster scan data and the optical metrology it is possible to determine that there is no insensitive region at the edge of the active edge pixels, with the edge-most row of pixels only dropping to 60% of the efficiency of the pixels in the centre of the device. The lower efficiencies and distortions observed at the edge pixels match those seen in [1] for this type of device.

A simple flat field correction is all that is required to adjust for the under and over efficient pixels in the extreme edge regions. A silicon powder diffraction measurement was taken to demonstrate the utility of the sensor array in one of its most common proposed applications.

With the manual placement of the detector assemblies in the array a gap down to 16um was achieved, with a more controlled gluing procedure an even closer and uniform packing is achievable.

These edgeless sensors are well suited to being arranged in a large array to replace a traditional large single sensor with multiple ASICs. This is particularly the case in Photon Science applications where a large close packed array of flat tiled detectors is required.

To reduce the gap between chips further a more controlled gluing method using vacuum clamping could be developed rather easily, and this will be the next subject investigated in this project.

Further analysis of this data set to match the corner pixel shapes to the expected field lines could be carried out. However, as this sensor series has already been superseded by an improved version with less distortions in the edge pixels this, work has not been prioritised.

### 8) Acknowledgments